# Reducing Errors in Excel Models with Component-Based Software Engineering

Craig Hatmaker

*Craig_Hatmaker@Yahoo.Com*

**ABSTRACT**

*Model errors are pervasive and can be catastrophic. We can reduce model errors and time to market by applying Component-Based Software Engineering (CBSE) concepts to Excel models. CBSE assembles solutions from pre-built, pre-tested components rather than written from formulas. This is made possible by the introduction of LAMBDA. LAMBDA is an Excel function that creates functions from Excel's formulas. CBSE-compliant LAMBDA functions can be reused in any project just like any Excel function. They also look exactly like Excel's native functions such as SUM(). This makes it possible for even junior modelers to leverage CBSE-compliant LAMBDAs to develop models quicker with fewer errors.*

## 1   INTRODUCTION

This paper opens by discussing Excel as a programming language that makes applying CBSE concepts to Excel possible. Section 3 discusses CBSE's value to both modelers and end users. Section 4 explains what qualifies a component as CBSE compliant. Section 5 discusses how component creators can make their components discoverable and how component consumers can include them in their projects. Section 6 walks us through assembling a real-world project. Section 7 summarizes this paper and is followed by references. An appendix provides answers to questions I have encountered and questions I expect from the modeling community.

## 2   EXCEL AS A PROGRAMMING LANGUAGE

This section discusses Excel as a programming language. Some of this paper's references use computer science terms. These terms may be unfamiliar to many Excel professionals so the table below may be helpful.

| CBSE Term | Excel Equivalent |
| --- | --- |
| Component | An Excel workbook, worksheet, chart, or function whether native, user-defined or from add-ins. |
| Component Consumer | Someone who uses components |
| Component Creator | Someone who develops CBSE compliant LAMBDAs. |
| Component Interface | A function's parameters (aka function arguments) |
| Defect Density | Errors per formula cell |
| Developer, Programmer | Modeler |
| Line of Code | A formula in a cell |
| Programming Language | Excels functions, operators, references, and formula syntax rules |
| System, Application | Completed model |

Excel is the world's most widely used programming language [Microsoft 2021]. Though long accepted as a programming language, Excel lacked features that would qualify it as a Turing-complete programming language until the introduction of LAMBDA. As a Turing-complete programming language, Excel users can, in principle, write any computation. [Microsoft 2023].

LAMBDA was announced in December 2020 and made generally available in February 2022. It is an Excel function with which users can define new functions from Excel formulas. Before LAMBDA, creating user-defined functions (UDFs) required knowledge of programming languages like VBA. Because LAMBDA's programming language is normal Excel formulas, anyone who can write Excel formulas can also create UDFs. The LAMBDA programming language includes all Excel functions but also features unique to LAMBDA as an Excel function that greatly expands Excel's capabilities.

**Recursion**

Recursion is a feature LAMBDA shares with most programming languages but not with any other Excel function. Recursion is the ability for a function to call itself and adds the ability to process hierarchical data structures to Excel. A simple hierarchical data structure example is the Fibonacci sequence. The Fibonacci sequence starts with 0 and 1. The next number in the sequence is the sum of its two predecessors; thus, the third number is 1 (0 + 1), the fourth number is 2 (1 + 1) and the third number is 3 (1 + 2).

To calculate the $n^{th}$ number, we can write an elegant recursive formula like the one in Figure 1.

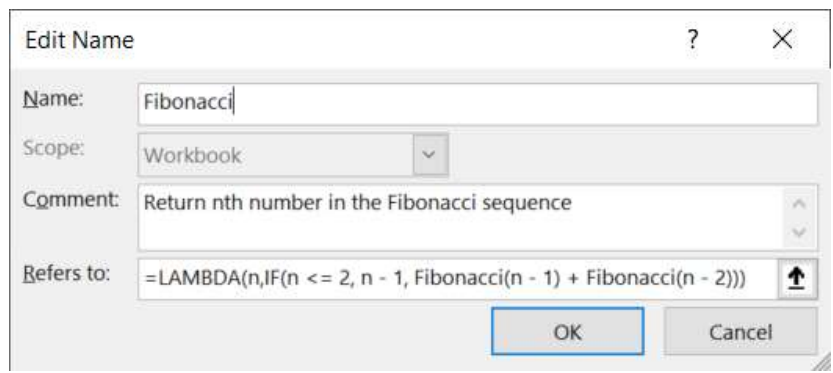

Figure 1 Recursive LAMBDA displayed in Excel's Name Manager

To use it in a cell to find the 9$^{th}$ number in the Fibonacci sequence we would write =Fibonacci(9). The function is intuitive, and the formula is compact.

Recursion can be simulated by enabling circular equations via Excel's menu path File > Options > Formulas > Enable iterative calculation and setting the maximum iterations. Unfortunately, this enables all circular equations, even ones we did not intend to be circular. Recursive LAMBDAs do not need this option checked and can limit recursion to the confines of a single LAMBDA. LAMBDA recursion has no impact on other formulas, and it does not permit LAMBDAs with external circular references to iterate.

**LAMBDA's Helper Functions**

LAMBDA has several helper functions unique to it including BYROW, BYCOL, MAKEARRAY, MAP, REDUCE, and SCAN. These make dynamic arrays work when they would not otherwise. Dynamic Arrays by themselves can reduce error potential by addressing inconsistent formulas which is a common spreadsheet error [O'Beirne 2010] [O'Beirne 2014] [Mkamanga K 2021]. With dynamic arrays, one formula entered into one cell can spill into tens of thousands of other cells which simply cannot be inconsistent. These functions with LAMBDA greatly expand how we can work with dynamic arrays to reduce errors and speed delivery.

Another function that can be used without LAMBDA, but which helps enormously is LET(). LET can mimic a feature shared by all programming languages like VBA known as local variables. Local variables can only be accessed within the function. This means they cannot interact with other functions even if those other functions have a variable of the same name. LET's version of local variables is locally defined named formulas. Locally defined named formulas are very similar to named formulas defined in Excel's Name Manager but as locally defined named formulas, they never appear in name manager. Another advantage of locally defined formulas is they can segment complex formulas into smaller, easier-to-understand, and easier-to-test sections with their intent conveyed by their name.

### 2.1 Section Summary

Excel is a programming language that everyone writing formulas uses. LAMBDA extends Excel's programming language to make it Turing-complete like other programming languages such as VBA. As a Turing-complete programming language, we can leverage the vast knowledge base for Turing-complete languages compiled over many decades such as CBSE.

## 3  COMPONENT-BASED SOFTWARE ENGINEERING VALUE PROPOSITION

Component-based software engineering (CBSE), also known as component-based development (CBD), emerged in the late '90s as an approach to software systems development based on reusing software components [Sommerville-2016]. Microsoft Excel is an example of a component-based application [Brown, 2000]. CBSE's value promise includes improved solution complexity management, increased productivity, reduced time to market, improved consistency, improved quality, and reduced maintenance costs, among other things. [Ivica 2001] [Brown [2000]. While CBSE provides many benefits, this paper concentrates on just two: reducing errors, and time to market.

### 3.1  Errors and Productivity

Model errors are pervasive [Panko, 2005] and can be catastrophic [O'Beirne 2023]. CBSE addresses errors by leveraging pre-tested reusable components. The theory is components designed for reuse will be tested more thoroughly before implementation [Mohaghegh 2004]. Additionally, errors will be exposed as a component is used in varied situations. Once exposed the function will be remediated and then placed back into production. This process repeats until no more errors surface. The monitor, fix, and repeat cycle is not unique to the software industry. It is also part of many manufacturing quality assurance best practices including CIP (Continual Improvement Process) [Wikipedia 2023-1], TQM (Total Quality Management) [Wikipedia 2023-2], and Six Sigma [Wikipedia 2023-2].

In the software industry, this theory has been tested by several researchers. Researchers from the MITRE corporation measured the impact of code reuse and defect density. They determined high reuse reduced errors by as much as 50% [Agresti 1992]. Researchers in Norway conducted similar research and also concluded that reused components resulted in nearly 50% fewer defects [Mohaghegh 2004]. A study in 1995 found an even stronger reduction in errors [Melo 1995].

I have found this to be true in my own experience as a developer, I have dramatically reduced errors and my time to deliver through code reuse. Solutions that I estimate would have taken me many months to code, took me weeks by assembling components from purchased libraries. Researchers have measured the correlation between code reuse and productivity and found timesaving not just from reusing components instead of coding them, but also in a reduction in testing and rework due to the higher quality of pre-tested components. [Melo 1995].

### 3.2  Additional benefits

LAMBDA components can codify business logic. Business logic is the expression of business rules in code. For example, if a company has a rule on which DCF (Discount Cash Flow) method they use and which variation of WACC (weighted average cost of capital) they use to calculate DCF, that formula can be placed into a reusable component. Then, the company can make that rule accessible to its modelers and allow only that component to be used in their projects.

### 3.3  Section Summary

CBSE has been proven to reduce errors and speed delivery. It can also be used to enforce business rules in models.

## 4  WHAT IS A CBSE COMPONENT?

The previous section explains why CBSE is important to modeling. This section explains what CBSE is within the Excel context.

There are several definitions for a CBSE component that I will interpret and consolidate.

A CBSE component is:
- designed for reuse without modification
- within the Excel application
- providing non-trivial functionality
- which is accessible through its function parameters/arguments.
- It is deployed as a self-contained independent object
- Ideally, it should be well-documented,
- provide user feedback and be rigorously tested
  [Brown, 2000] [Gloag 2018] [Sarvani 2015] [Szyperski 2002]

Excel's native functions provide over 500 examples. Each function is designed to be reused without modification many times within the same Excel workbook and other Excel workbooks. Each function provides non-trivial functionality that we cannot modify, and that we can only interact with through their parameters (arguments). They are all well documented through online help and rigorously tested.

### 4.1  Creating CBSE-Compliant LAMBDAs

In the CBSE context, there are two classes of users, component creators and component consumers. A component creator is someone who develops CBSE-compliant LAMBDAs. Component creators are typically highly skilled though that is not always a requirement. Component consumers are those who use components without having to know how the components work. This section is for component creators.

One CBSE requirement is the function must be non-trivial. While it is possible to write trivial LAMBDAs, the cost to create reusable components meeting all requirements can only be justified through reuse [Frakes, 1996]. So let us look at the extra work required to make LAMBDA function reusable.

The LAMBDA function has three sections: parameter names, the calculation, and parameter

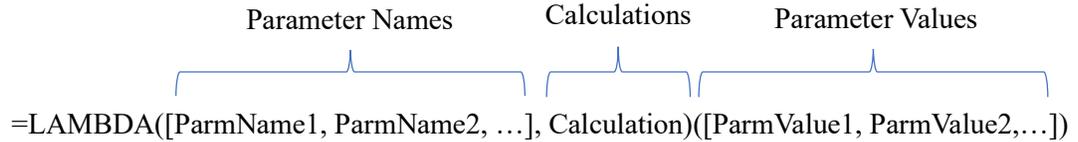

Figure 1 LAMBDA Sections

values.

How component creators use these sections determines if a LAMBDA can be reused everywhere. To be reused everywhere LAMBDAs must be self-contained [Szyperski 2002]. To be self-contained, the LAMBDA calculation section must not reference cells directly as that would make them dependent on the host worksheet. Table 4 provides examples of the CBSE-compliant way to incorporate cell references (row 2) and the non-compliant way (row 3). The compliant way passes cells as parameters so we can pass these values without modifying the LAMBDA's calculation section. The non-compliant LAMBDA requires the values in cells A2 and B2. The non-compliant formula will break if we move the values to any other cells.

|   | A | B | C | D |
|---|---|---|---|---|
| 1 | Parm 1 | Parm 2 | Result | Formula Text |
| 2 | 1 | 2 | 3 | =LAMBDA(Parm1,Parm2, Parm1 + Parm2)(A2,B2) |
| 3 |   |   | 3 | =LAMBDA(Parm1,Parm2, A2 + B2)(3, 4) |

Table 1 CBSE compliant (green) and non-compliant (red) LAMBDAs

This rule also applies to LAMBDA's helper functions: BYROW(), BYCOL(), MAKEARRAY(), MAP(), REDUCE(), and SCAN(). Table 5 provides examples of the CBSE-compliant way to incorporate cell references (row 6) and the non-compliant way (row 2). The compliant way passes cells as parameters which requires wrapping these formulas in an outer LAMBDA. This adds a little more work.

|   | A | B | C | D |
|---|---|---|---|---|
| 1 |   | Parm 1 | Result | Formula Text |
| 2 |   | 1 | 2 1,2 | =BYROW( A2:B4, LAMBDA(Row, TEXTJOIN( ",", TRUE, Row))) |
| 3 |   | 3 | 4 3,4 |   |
| 4 |   | 5 | 6 5,6 |   |
| 5 |   |   |   |   |
| 6 |   | 1 | 2 1,2 | =LAMBDA(Array, BYROW( Array, LAMBDA(Row, TEXTJOIN( ",", TRUE, Row))))(A6:B8) |
| 7 |   | 3 | 4 3,4 |   |

Table 2 CBSE compliant (green) and non-compliant (red) LAMBDAs with helper functions

LET() is not unique to LAMBDA but to create CBSE-compliant LAMBDAs that use LET, LET must not contain references to cells or named formulas. The diagram below shows the LET function in row 2 directly accessing dynamic arrays A2# and B2#. This is not CBSE-compliant. To be CBSE compliant we can wrap the LET() in a LAMBDA and receive the dynamic arrays via the LAMBDA's parameters.

|   | Parm 1 |   | Result | Formula Text |
|---|---|---|---|---|
| 1 |   |   |   |   |
| 2 | 1 | 2 | 3 | =LET( |
| 3 | 2 | 3 | 5 |     Array, HSTACK(A2#,B2#), |
| 4 | 3 | 4 | 7 |     BYROW(Array, LAMBDA(Row, SUM(Row))) |
| 5 | 4 | 5 | 9 | ) |
| 6 | 5 | 6 | 11 |   |
| 7 |   |   |   |   |
| 8 | 1 | 2 | 3 | =LAMBDA(Column1,Column2, |
| 9 | 2 | 3 | 5 |     LET(Array,HSTACK(Column1, Column2), |
| 10 | 3 | 4 | 7 |         BYROW(Array, LAMBDA(Row,SUM(Row))) |
| 11 | 4 | 5 | 9 |     ) |
| 12 | 5 | 6 | 11 | )(A8#,B8#) |

Table 3 CBSE compliant (green) and non-compliant (red) LAMBDAs with LET

### 4.2 Documenting Components

CBSE-compliant LAMBDAs require documentation that forms the component specification. The documentation must include a description of the services it provides, a list of allowable inputs with their constraints, a description of outputs, and any conditions for which the component will and will not function. [Brown 2000] [Szyperski 2002]. Below is an example

```
/*  FUNCTION NAME:   SumColumnsλ                                           ── Component Name
    DESCRIPTION:*//**Totals an array's columns*/
/*  REVISIONS:       Date           Developer        Description
                     Mar 06 2023    Craig Hatmaker   Original Development
                     Apr 10 2023    Craig Hatmaker   Added Help
*/                                                                         ── Revision log

SumColumnsλ = LAMBDA(
//  Parameter Declarations
    [Array],
//  Procedure
    LET(Help,        TRIM(TEXTSPLIT(
                     "DESCRIPTION:   →Creates totals for each column in an array.¶" &
                     "VERSION:       →Apr 10 2023¶" &
                     "PARAMETERS:→¶" &
                     "Array          →(Required) A two dimensional array of values.¶" &
                     "→¶" &
                     "EXAMPLES:→¶" &
                     "Result         →Formula¶" &
                     "15,25,35,45    →=SumColumnsλ({1,2,3,4;4,3,2,1;10,20,30,40})" ,
                     "→", "¶" )
                     ),
//  Check inputs
        Array,       IF(OR(ISOMITTED(Array), Array=""), #Value!, Array),
//  Procedure
        Result,      BYCOL(Array, LAMBDA(Column, SUM(Column))),
//  Handle Error
        Error,       MAX(IsError(Result)+1),
//  Return Result
        Choose(Error, Result, Help)
    )
);
```

*Inline help shows when no parameters are entered, or the LAMBDA fails. See help output below source.*

|   | A | B |
|---|---|---|
| 5 | DESCRIPTION: | Creates totals for each of an array's columns. |
| 6 | VERSION: | Apr 10 2023 |
| 7 | PARAMETERS: | |
| 8 | Array | (Required) A two dimensional array/range containing values to be summed. |
| 9 | | |
| 10 | EXAMPLES: | |
| 11 | Result | Formula |
| 12 | 15,25,35,45 | =SumColumnsλ({1,2,3,4;4,3,2,1;10,20,30,40}) |

- Description of services provided.
- Allowable inputs.
- Example usage

Figure 2 Example CBSE compliant documentation for LAMBDAs

of a LAMBDA source with documentation meeting these requirements as displayed in the Advanced Formula Environment.

### 4.3 Section Summary

Converting good LAMBDAs to CBSE-compliant LAMBDAs requires a little extra work which centers on moving all cell and named formula references from the LAMBDA's calculations section, helper formulas, and LET functions to the LAMBDAs parameters section. Good documentation is also required. Adding features like feedback is ideal.

## 5 USING CBSE-COMPLIANT LAMBDAS

Component creators need to store their components in a repository they can write to, and component consumers can read from. For components codifying business logic, the repository can be a network directory that only members of the company can access. If the component creator wishes to freely distribute their LAMBDAs, a free repository is a public GitHub Gist (https://Gist.GitHub.com) which Microsoft's free Advanced Formula Environment (AFE) add-in for Excel integrates with to import LAMBDAs directly into Excel. AFE also integrates with private GitHub Gist which component creators can use for limiting access.

Component consumers need the means to find components. At the time of this writing, there are very few CBSE-compliant LAMBDAs available, and none can be found by Google search. To find them, component consumers can go to Gist.GitHub.com and use its search facility with the search string "CBSE compliant LAMBDA". Gists designated as "public" can be discovered this way. Gists designated as "private' can only be accessed by URL. An option to find private Gists exists at Eloquens.com which has a market for LAMBDAs.

Once a desired component is found and its URL obtained, component consumers can use Microsoft's free Excel add-in, Advanced Formula Environment (AFE) [William 2023]. AFE is designed to import LAMBDA Gists referred to as modules by AFE. AFE imports modules, directly into Excel. To avoid the potential of imported LAMBDAs having the same name as other items in Name Manager, AFE asks consumers to provide a short ID for the module which AFE applies as a prefix to each LAMBDA from that Gist. AFE stores the imported LAMBDAs in the workbook's XML and makes them available within that workbook like any other Excel function.

Gists can be updated either to correct a problem or improve functionality. To preserve backward compatibility, previous version interfaces (parameters) must not change but new parameters can be added to the end of the interface definition as optional. When changed, GitHub logs the changes in detail and increments the version number. Component Consumers can check their version against what is available in GitHub to determine if they have the most current version, and if not, what the differences are.

## 5.1 Section Summary

Component creators can store components freely in GitHub Gists. Component consumers can find public components using the GitHub Gists search facility. Once found, component consumers can use AFE's import feature to load modules directly into Excel where they can be used like any other Excel function. Example

The following example is for building and managing a departmental budget. Notable features include:

1. It can handle any number of accounts, budget items, actual expenditures, and periods without modeler intervention.
2. No refresh is required (except to download actuals if using Power Query).
3. Calculations can be assembled in 6 minutes.

It uses two LAMBDA libraries:

1. Excel CBSE Compliant LAMBDA Dates (module prefix used BXD):
   https://gist.github.com/CHatmaker/3e1708888ec2bd1cde2ec9d002dc459b
2. Excel CBSE Compliant LAMBDAs for Reporting with Arrays (module prefix used BXR):
   https://gist.github.com/CHatmaker/cc0e8975d30b40734641a06dbae02143

The template with worksheets and modules but no calculations can be obtained here:
https://www.dropbox.com/s/8wtwexl8sarkdb2/CBSE%20Budget%20Template%20Blank%20with%20LAMBDAs.xlsx?dl=1

The completed example can be downloaded from here:
https://www.dropbox.com/s/j82p9v6vbmo5q8z/CBSE%20Budget%20Template%20Completed.xlsx?dl=0

A video showing the end-user experience of this model is available here:
https://youtu.be/uEXd8h1i3LM

A video showing assembling this model is available here:
https://youtu.be/ERzF2uw4cX4

The add-in I used in to simplify naming dynamic arrays is available here:
https://www.eloquens.com/tool/ELyYf0m4/finance/excel-tips/bxl-dan-free-create-dynamic-array-names-and-more

## 5.2 Worksheet Inputs

We start with the user inputs worksheet which has sample data already loaded. On the left are the model's properties which include the *Start* date, *End* date (blank defaults to one year), and reporting period *Interval*. These are named cells. On the right are budgeted items. Each item has an *Account*, amount (*Ext.Amt.*), *Schedule* code, and start date (*First Date*). Some items have an end date (*Last Date*). The schedule code designates whether a budget item repeats at regular intervals which include daily, weekly, biweekly, monthly, quarterly, semiannually, annually, or not at all.

## 5.3 Worksheet ItemSchedule

Our project's calculations start by scheduling all budget item amounts on worksheet *ItemSchedule*.

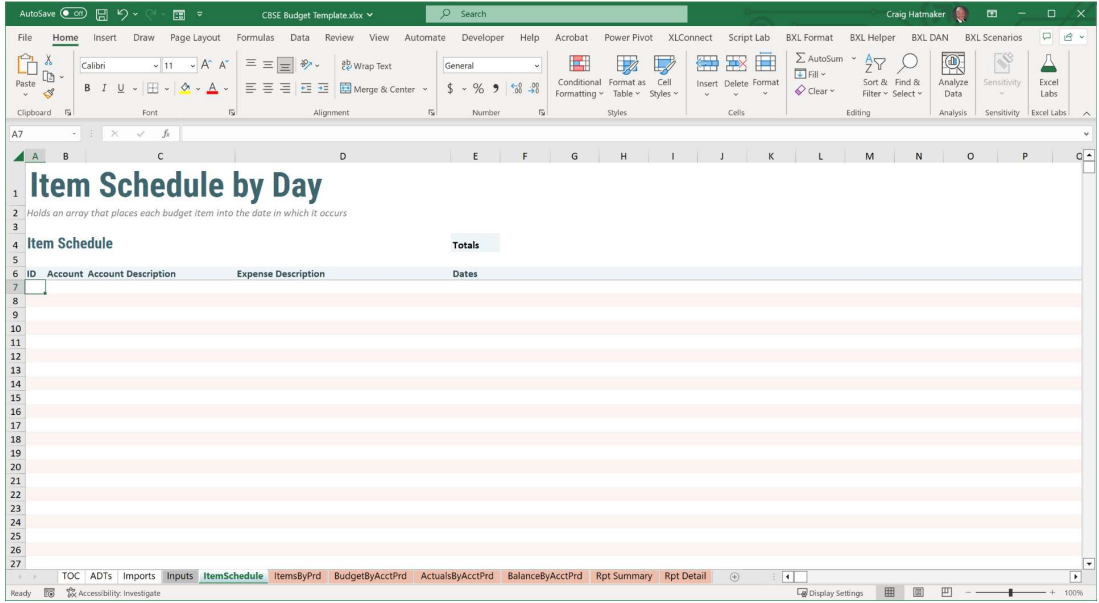

For this, we will use components related to scheduling and dates, and components related to reporting that perform grouping and summing functions. The components we need are already loaded into the template. I named the reporting components' module *BXR*. I named the date components' module *BXD*. We can see these components using the AFE. See the Appendix for answers on how to load and access the AFE.

These two modules contain an *Aboutλ* component. If we enter =BXD.Aboutλ into a cell we get information about the dates module and a component listing. See below.

| About: | CBSE compliant LAMBDAs dealing with dates. Suggested module name: BXD |
|---|---|
| Version: | Apr 14 2023 |
| Gist URL: | https://gist.github.com/CHatmaker/cc0e8975d30b40734641a06dbae02143 |
| Website: | https://sites.google.com/site/beyondexcel/home/excel-library/ |

| Function | Description |
|---|---|
| Aboutλ | Produces this table |
| Periodsλ | Determine the number of periods from date1 to date 2 inclusive |
| CreateStartDatesλ | Creates a horizontal list of start dates for a timeline |
| CreateEndDatesλ | Creates a horizontal list of end dates for a timeline |
| IsOccurrenceDateλ | Determine if a date passed is when a potentially repeating event happens |
| IsBetweenλ | Determine if a value is between a lower and upper limit |
| CountMonthDOWλ | Count instances in a month for a specific day of the week |
| OverLapDaysλ | Return how many days overlap two period ranges. |
| ScheduleValuesλ | Schedules values in a timeline from a schedule in a table. |
| RunningTotalλ | Creates a running total for a vector array |
| PeriodLableλ | Creates a lable for a date based on period interval |

The components we need from module *BXD* (dates) module are: *BXD.CreateStartDatesλ()* which creates a timeline, and *BXD.IsOccurrenceDateλ()* which determines if a date in the timeline is one upon which a budget item occurs. Below shows these components in their locations.

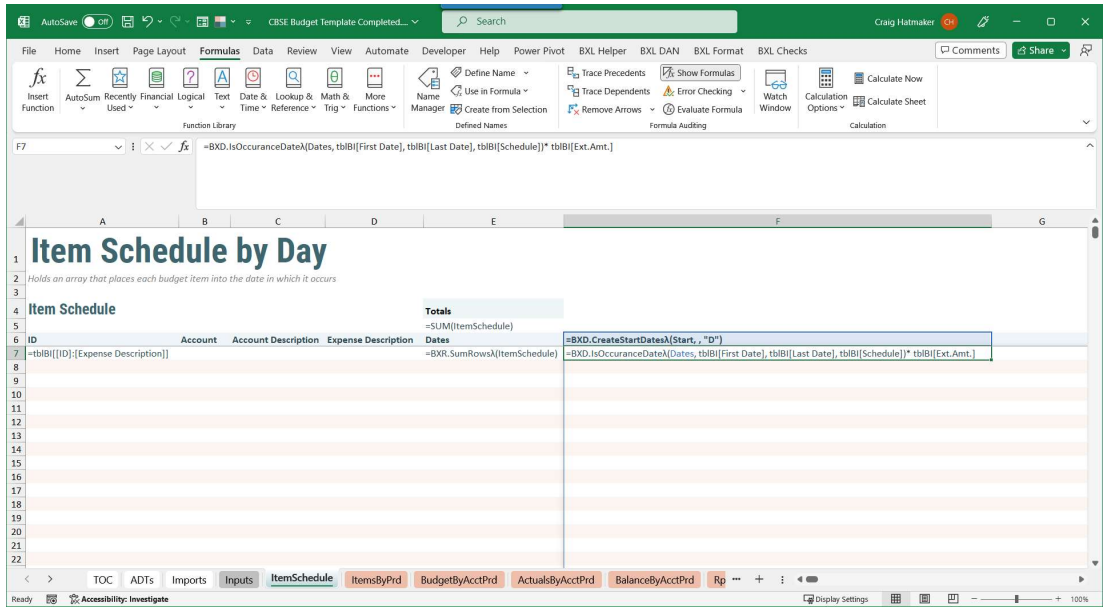

Before we turn our attention to the components, let us bring in the labels from our budget items table. Enter this formula in cell A7: =tblBI[[ID]:[Expense Description]]. At this point we should see an array of the *ID* through *Expense Description* columns from *tblBI* (Budget Items table on the *Inputs* worksheet).

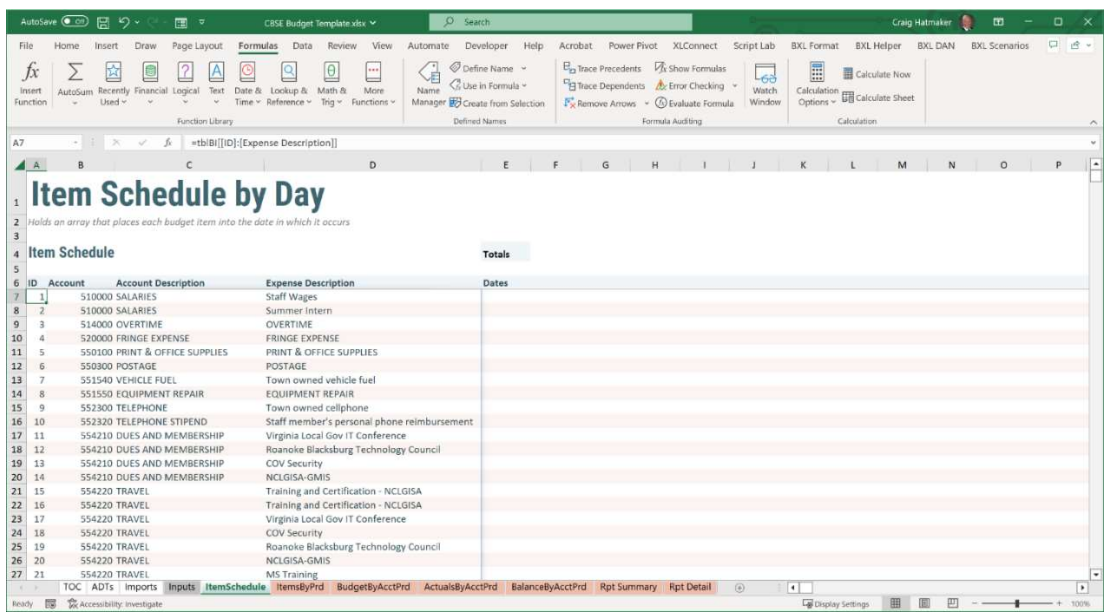

These labels are not part of our calculations. They are there just to help us relate calculated amounts to their associated budget item entries.

Let us now turn our attention to components. Because these are CBSE components, they include user feedback. One form of feedback is inline help. If we go to a blank worksheet and type into any cell: =BXD.CreateStartDatesλ(), the inline help spills into empty cells.

| DESCRIPTION: | Creates a horizontal timeline or period start dates. |
|---|---|
| | |
| PARAMETERS: | |
| StartDate | (Required) First date in timeline. |
| EndDate | (Optional) Last date in timeline. Defaults to 1 year after start. |
| Interval | (Optional) Difference in Days, Weeks, Months (default), Quarters, or Years |
| | |
| REQUISITES: | Requires Periodsλ() from this LAMBDA's source module |
| | |
| EXAMPLES: | |
| Result | Formula |
| 1 yr in Months | =CreateStartDatesλ("1/1/2023") |
| 1 yr in Days | =CreateStartDatesλ("1/1/2023", , "D") |
| 2 yrs in Qtrs | =CreateStartDatesλ("1/1/2023","12/31/2024", "Q") |

**NOTE!** We do not need to enter the lambda symbol (λ), nor the module prefix. I generally just start typing and by the time I reach *CreateS* Excel's intellisense autocompletes the formula for me and I can just hit ENTER.

Let us create a timeline. On our *ItemSchedule* worksheet enter this formula in cell F6: =BXD.CreateStartDatesλ(Start, , "D").

- *Start* is the named range on the *Inputs* worksheet which the user can use to designate when their budget year starts. For rolling budgets, this can be changed monthly and the model will automatically adjust.

- *End* date is omitted and will default to one year.

- "D" tells our component to set the timeline interval to daily.

At this point, we see a daily timeline extending out one year.

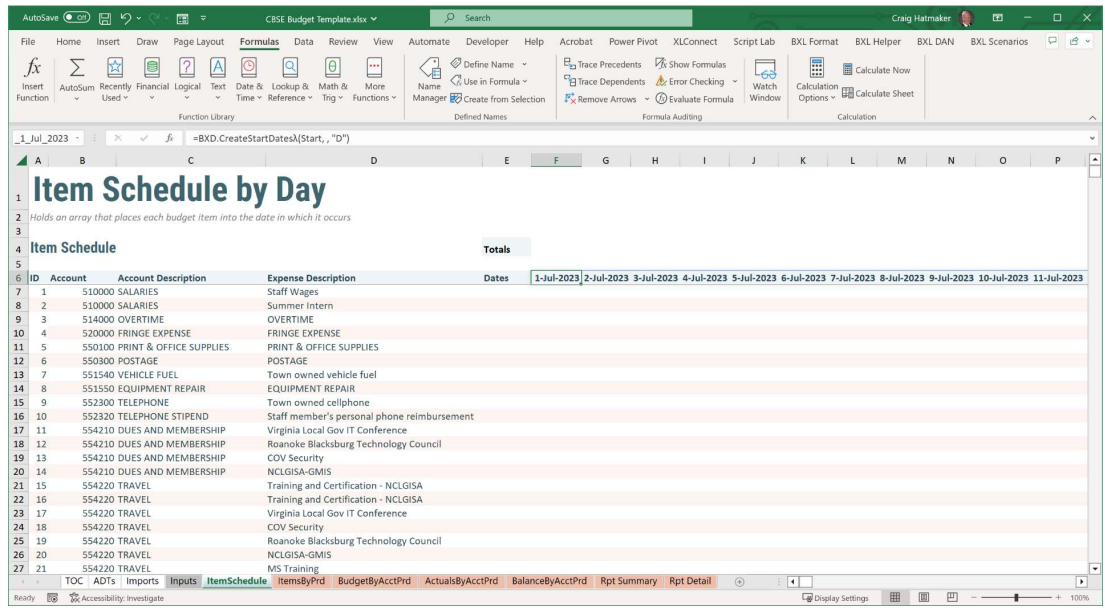

Let us name the timeline by selecting cell F6 and then use Excel ribbon tab Formulas > Define Name. Enter *Dates* in the Name: box and add a hashtag (pound symbol) in the Refers to: box after *=ItemSchedule!$F$6*. The hashtag makes the name refer to the array instead of just the cell. Click OK.

To schedule our budget amounts, enter this formula in cell F7: =BXD.IsOccurrenceDateλ(Dates, tblBI[First Date], tblBI[Last Date], tblBI[Schedule]).

- *Dates* is our timeline.

- *tblBI[First Date]* is the *First Date* column from our budget items table on our Inputs worksheet.

- *tblBI[Last Date]* is the *Last Date* column.

- *tblBI[Schedule]* is the *Schedule* column.

At this point, we see an array of TRUE and FALSE where TRUE is when a budget item's expense occurs.

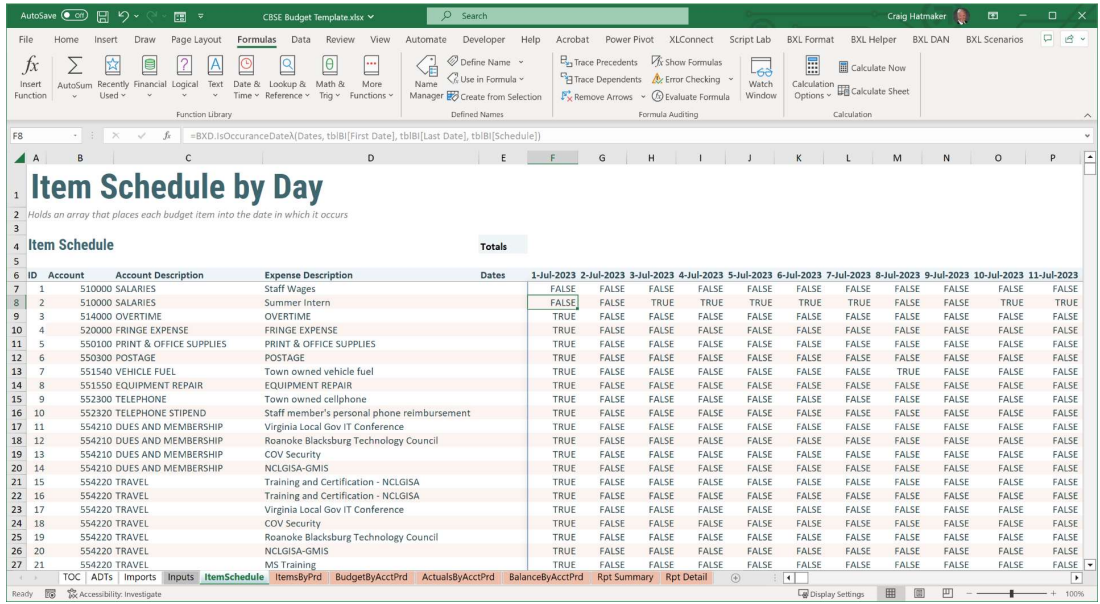

We can replace the TRUE and FALSE values with budget amounts by multiplying the formula by *tblBI[Ext.Amt.]*. Change the formula to this: *=BXD.IsOccurrenceDateλ(Dates, tblBI[First Date], tblBI[Last Date], tblBI[Schedule])* tblBI[Ext.Amt.]*. At this point, we see an array of budget amounts.

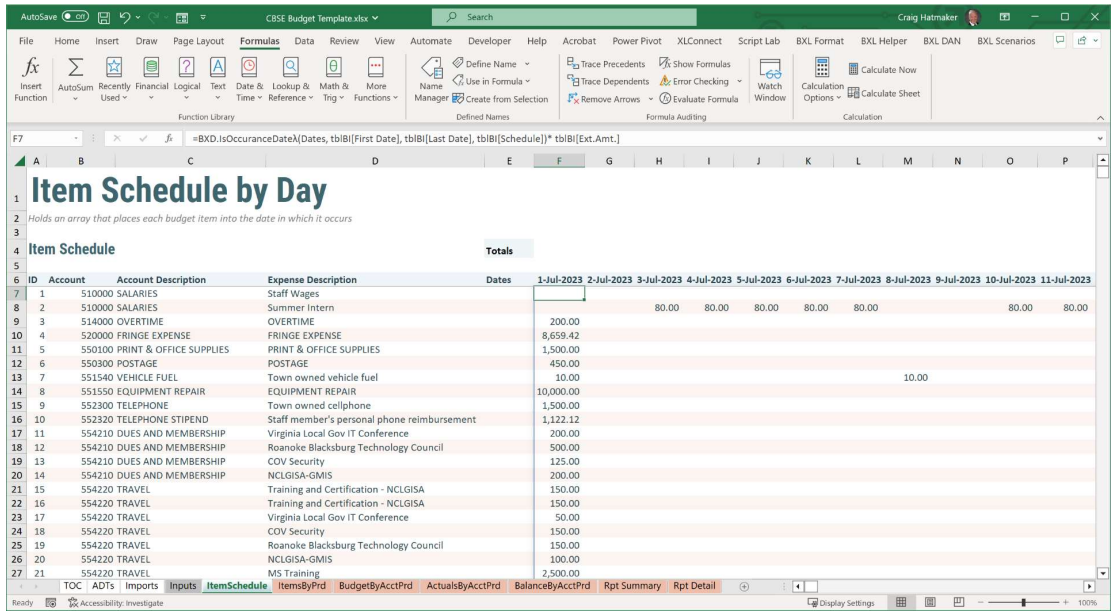

Let us name this array *ItemSchedule* by selecting cell F7 and then use Excel ribbon tab Formulas > Define Name. Enter *ItemSchedule* in the <u>N</u>ame: box and add a hashtag (pound symbol) in the <u>R</u>efers to: box after *=ItemSchedule!$F$7*. Click OK.

We total each row by entering the SumRowsλ component from module BXR. Enter this formula in cell E7: =BXR.SumRowsλ(ItemSchedule). At this point we see an array of totals for each budget item.

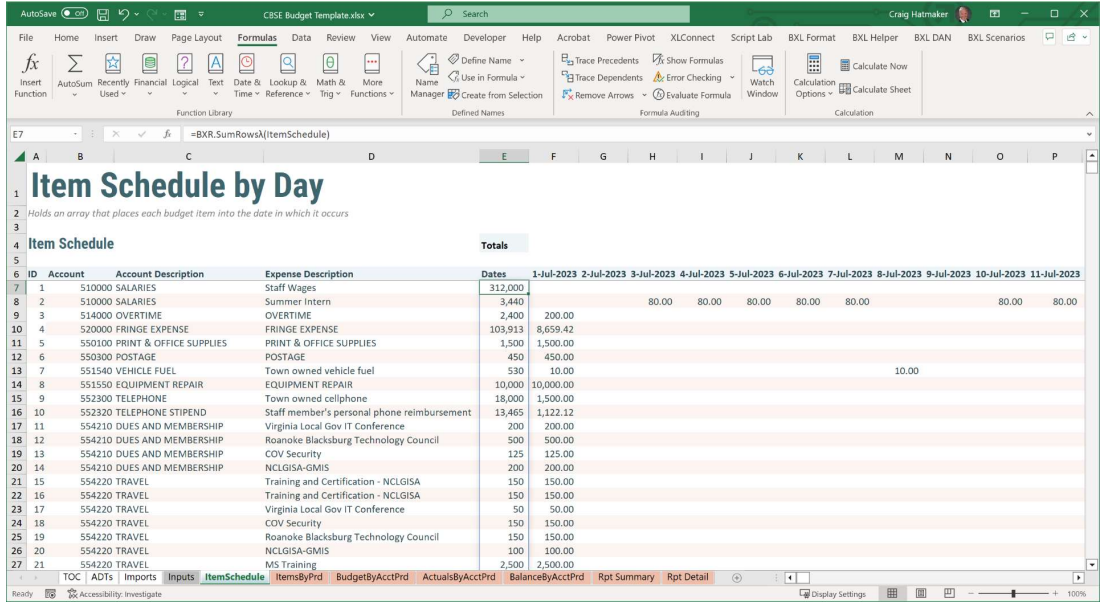

Lastly, we will total the entire array which we will use to compare against other sections of this model to confirm each budget section adds up to the same amount. In cell A5 enter this normal Excel formula: =SUM(ItemSchedule).

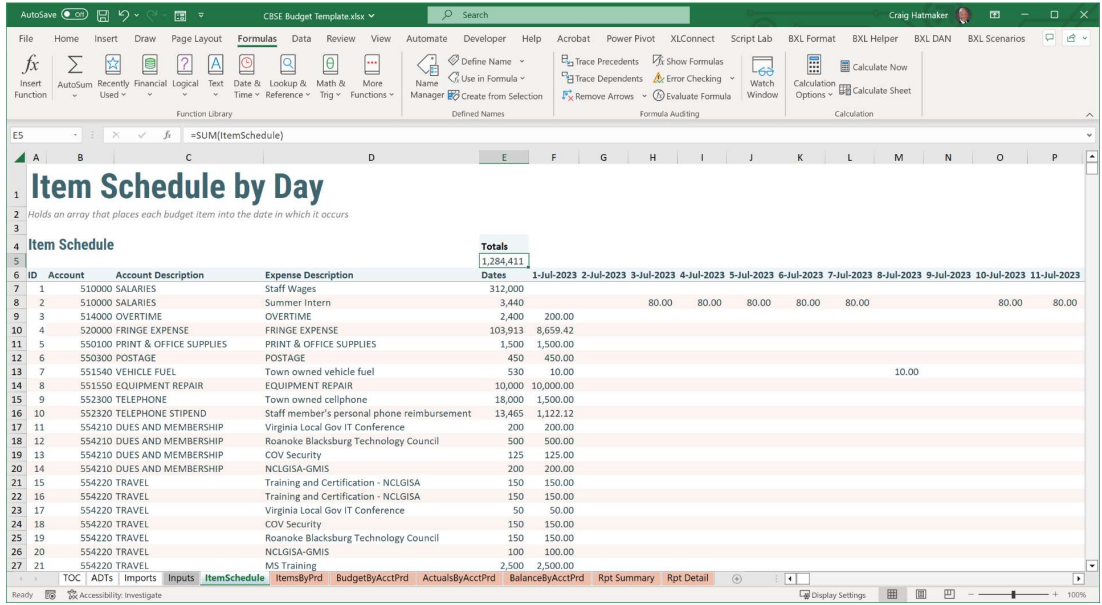

## 5.4 Worksheet ItemsByPrd

The next sections are assembled in a similar fashion. This section's purpose is to summarize budget items by reporting period (months in this case) instead of by day. It uses four components. Three are new: BXR.SumColumnsλ(); BXD.PeriodLabelλ(); and BXR.SumGroupsλ().

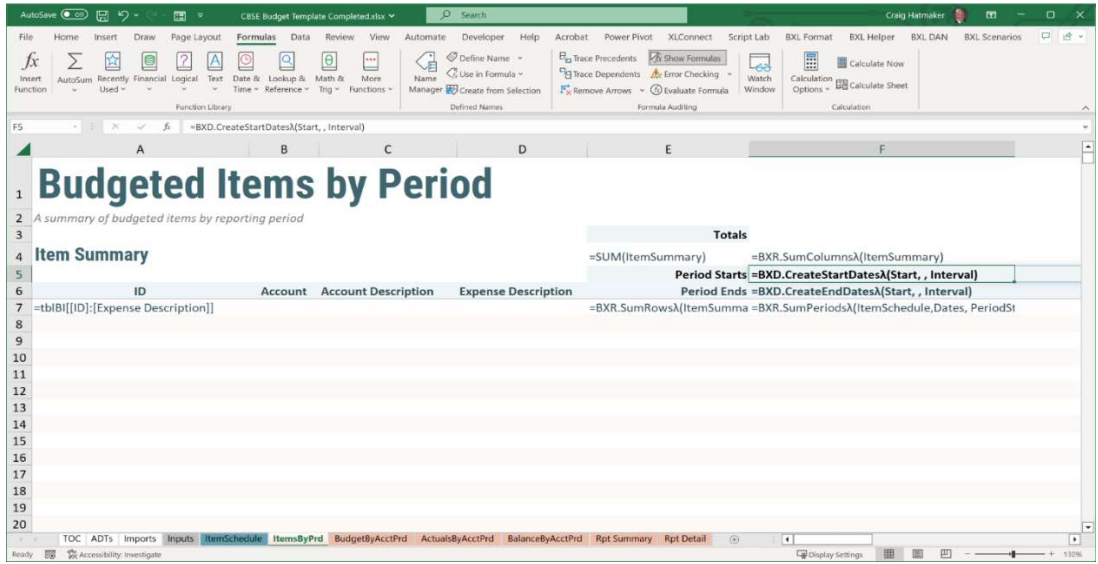

Below are the results after entering these formulas. I named the timelines: *PeriodStarts* and *PeriodEnds*. I named the main array: *ItemSummary*.

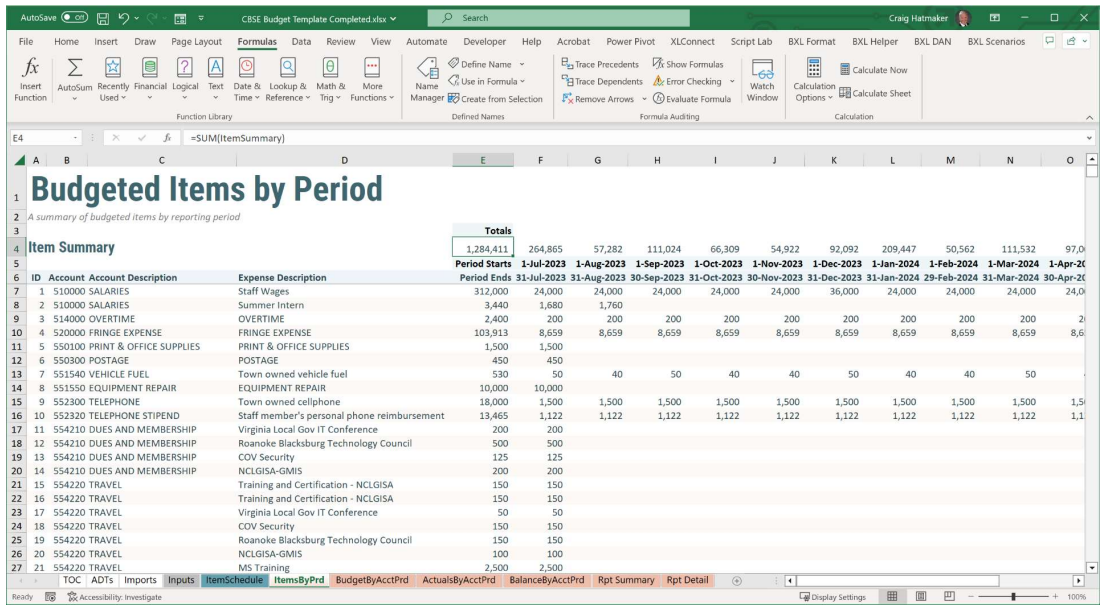

## 5.5 Worksheet BudgetByAcctPrd

This section's purpose is to summarize budget items by account instead of by item. It uses five components. Three are new: BXR.SumColumnsλ; BXD.PeriodLabelλ, and BXR.SumGroupsλ.

Below are the results after entering these formulas. I named the timelines: *Periods*. I named the main array: *BudgetSummary*.

## 5.6 Worksheet ActualsByAcctPrd

This section's purpose is to transform and summarize actual expenditures imported from our financial system. It uses three components. One is new: BXR.SumGroupsAndPeriodsλ. This component pivots the tabular data, then summarizes by account and period.

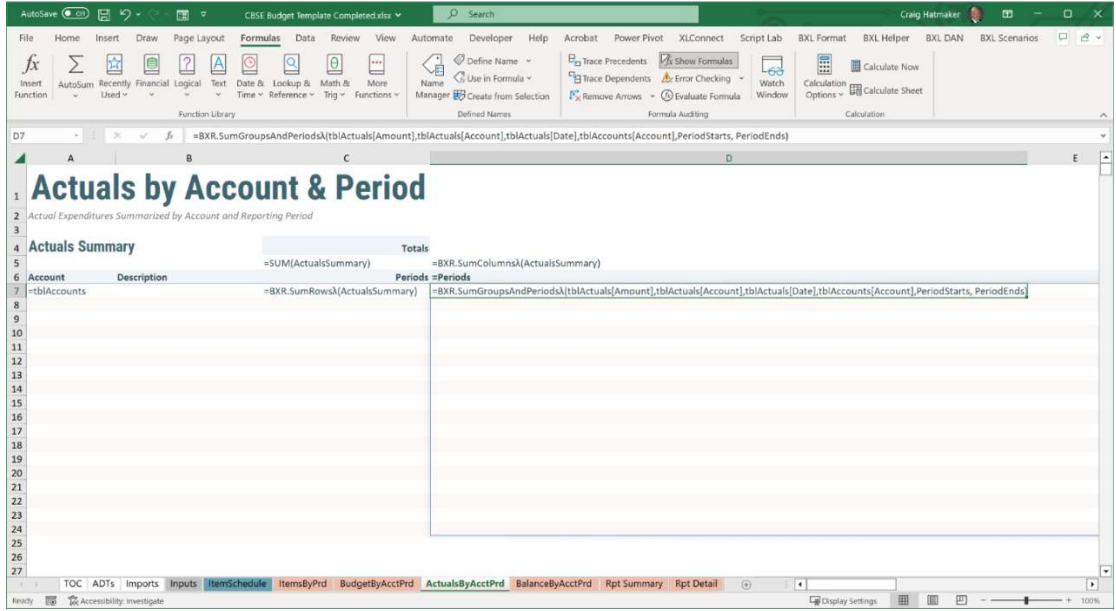

Below are the results after entering these formulas. I named the main array: *ActualsSummary*.

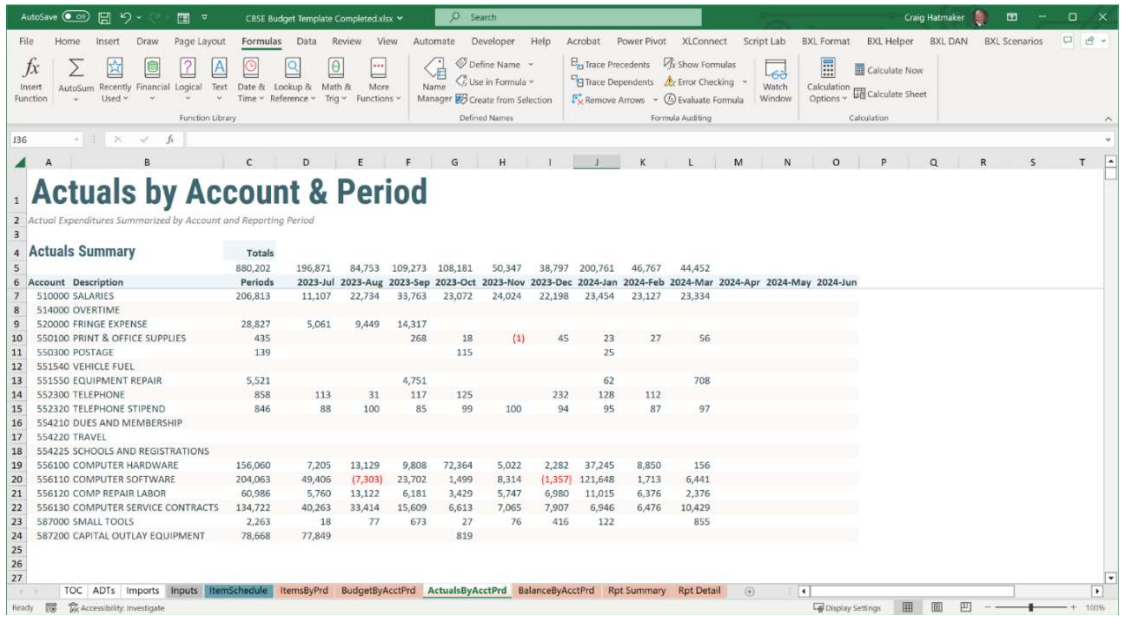

## 5.7 Worksheet BalancesByAcctPrd

This section's purpose is to calculate balances by subtracting actuals from budgeted. It uses two components. None are new.

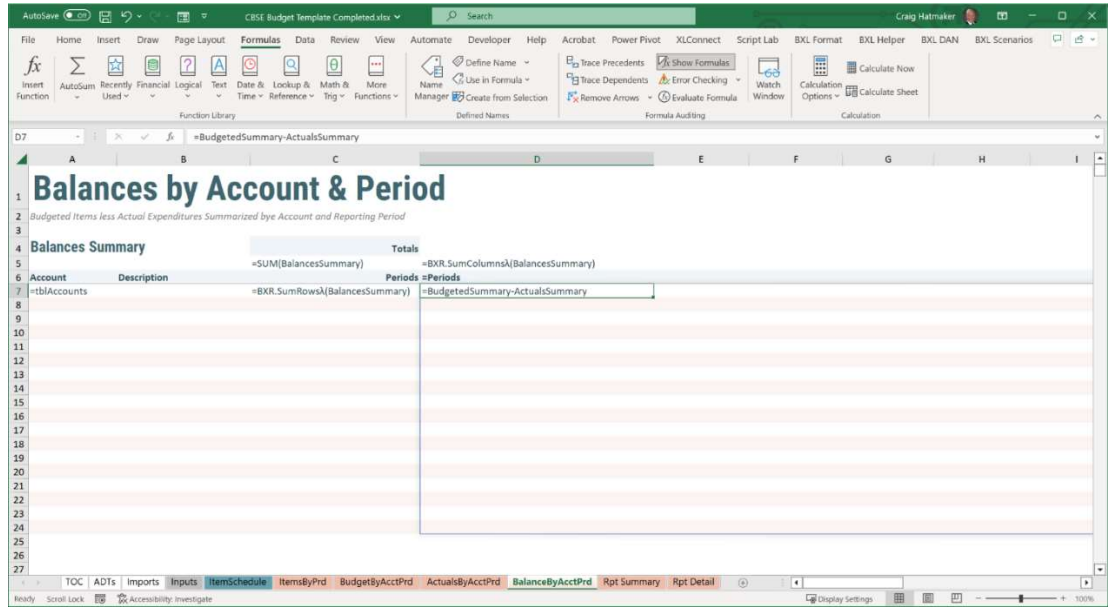

Below are the results after entering these formulas. I named the main array: *BalancesSummary*.

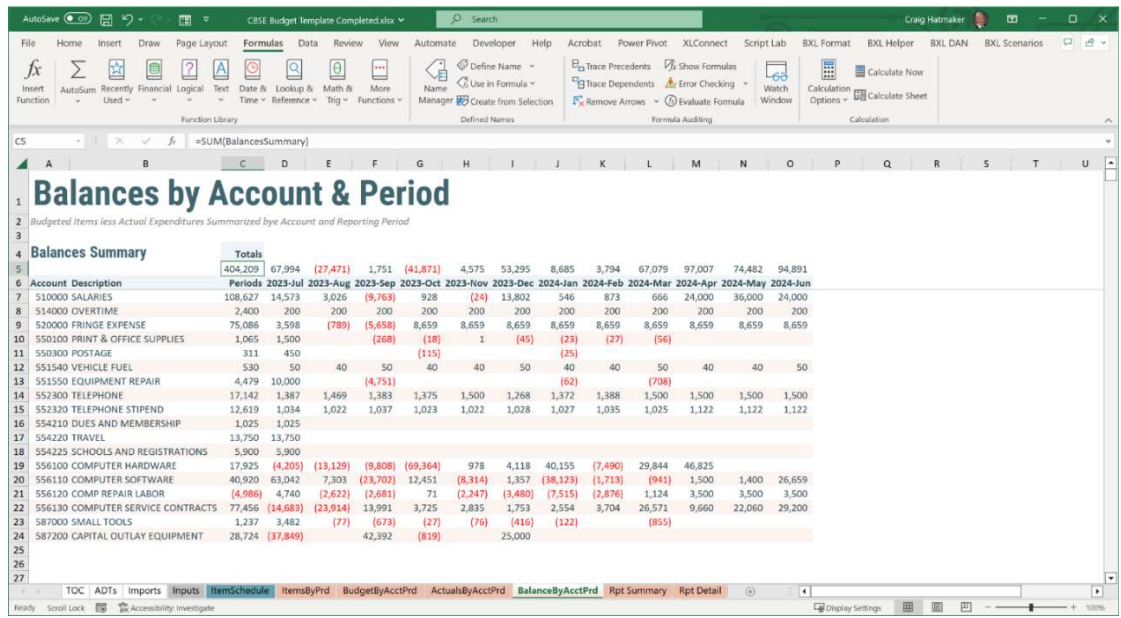

## 5.8 Worksheet Rpt Summary

This section's purpose is to present each budget account's budget totals, actual totals, period totals, and running totals. It uses one component: BXR.ReportGroupSummaryλ. The complete formula can be seen in the formula box.

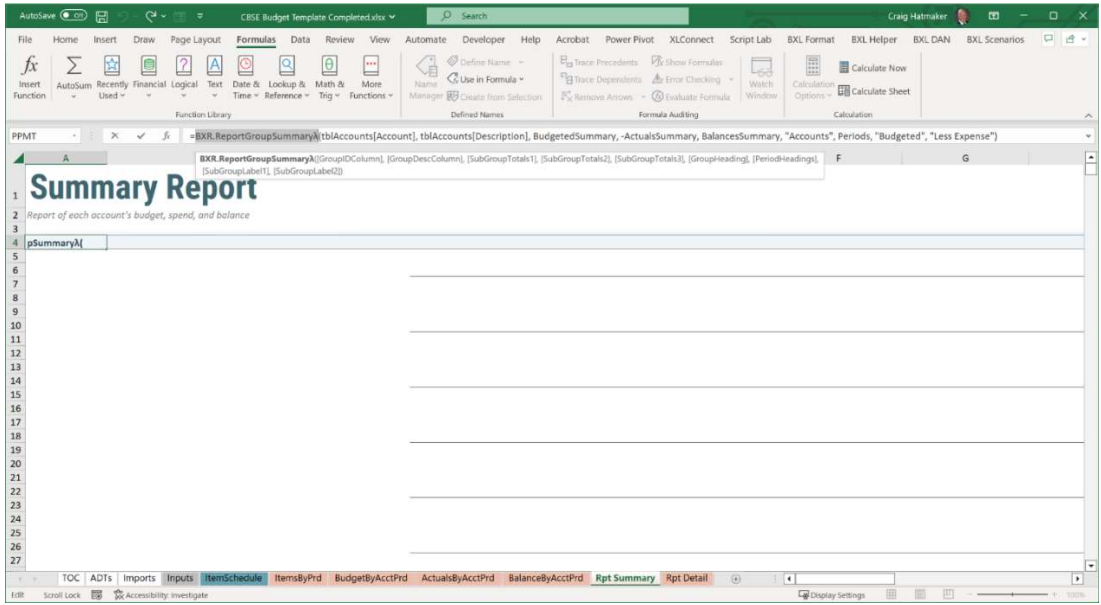

Below are the results after entering this formula.

## 5.9 Worksheet Rpt Detail

This section's purpose is to present each budget account's budget items, budget totals, actual totals, period totals, and running totals. It uses one component: BXR.ReportGroupDetailOffsetλ. The complete formula can be seen in the formula box.

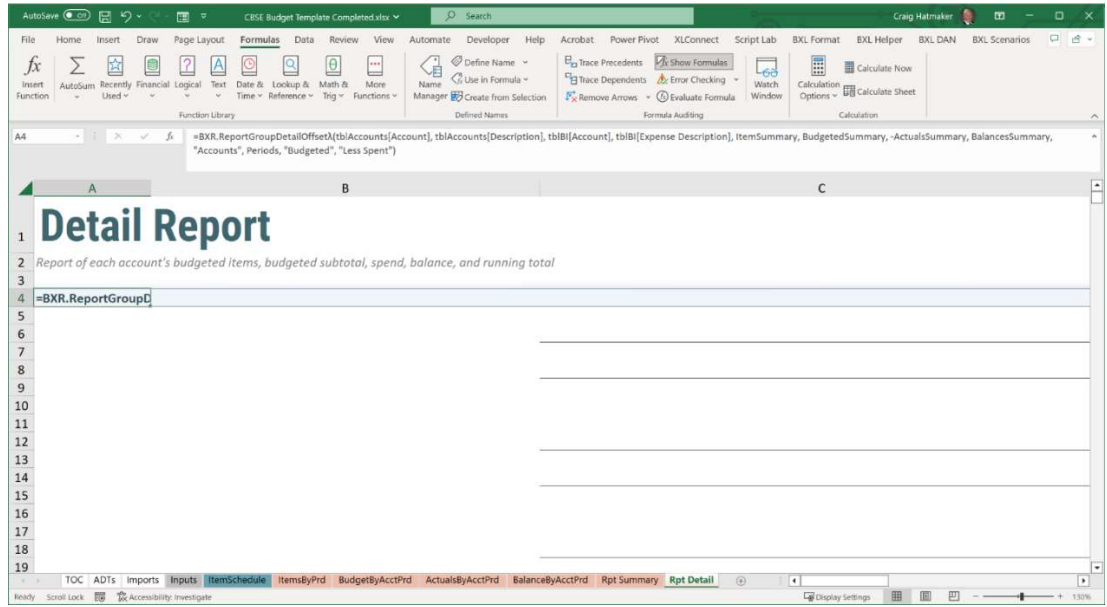

Below are the results after entering this formula.

## 5.10 Section Summary

This sample model provides a solution to a common problem: Departmental budgeting. It is flexible enough to satisfy account-level budgets, zero-based budgets, and rolling budgets. It can, without modeler intervention, accommodate any number of accounts, budget items, actual expenditures, and periods.

It uses less than a dozen distinct components that cover large model sections. Some components are specialized, such as the two reports. Some are general and can be used in many models and several times within the same model, such as the summing and timeline components. Because it uses so few formulas, it can be assembled in under 10 minutes.

## 6 SUMMARY

We can deliver projects far quicker by assembling models from components rather than writing formulas from scratch. We can deliver models that do things we do not know how to do by leveraging components built by experts. We can enforce corporate business rules by encapsulating business logic into pre-built, simple-to-use components. We can deliver models with fewer errors. This is why CBSE was developed decades ago which is now possible to implement in Excel with LAMBDA.

Creating complex LAMBDAs requires more skill; however, using CBSE components requires only remedial Excel skills [Microsoft 2021].

**APPENDIX**

**7    QUESTIONS AND ANSWERS**

Q: How do we get the Advanced Formula Environment (AFE) add-in?

A: From Excel's ribbon click **Insert** > **Get Add-Ins**. Search for *Excel Labs*. Click **Add** next to ***Excel Labs, a Microsoft Garage project***, then ***Continue***. The add-in is now part of Excel and is available to all workbooks.

Q: Where is the AFI add-in?

A: The add-in is located on Excel's **Home** tab under the **Excel Labs** icon on the far right. Click *Excel Labs* icon then *Advanced Formula Environment*.

Q: Do we have to distribute the Advanced Formula Environment (AFE) add-in with our workbooks?

A: No. AFE is only useful for importing and creating LAMBDAs. It is not needed to execute LAMBDAs.

Q: Where are LAMBDAs stored?

A: LAMBDA names, formulas, and descriptions are stored in Name Manager. LAMBDAs imported from GitHub Gist using AFE have their complete GitHub Gist source stored in the workbook's XML.

Q: Can I change an imported LAMBDA?

A: Yes. We can change LAMBDAs inside Name Manager, in AFE, in Excel, or even with Notepad.

Q: What happens to my workbook if a new version of a LAMBDA is released on GitHub Gist?

A: Nothing. The old version stays in our workbook. To get the updated version we must delete the old module and download the new version.

Q: Why would a LAMBDA creator come out with new versions?

A: They may have found a bug and fixed it, a more efficient formula, updated documentation, or added a new feature.

Q: If a new feature is added to a LAMBDA, can it break our workbooks?

A. Not if we don't import it and not if done correctly. As long as the original parameters remain in their positions, and the new features are accessed as optional parameters added to the end of the original parameters, our existing workbooks should be able to import the new versions with no issues.

Q: If a LAMBDA is updated, can we see what changed?

A. Yes if stored on GitHub Gist which has a very robust version control system.

Q: Are LAMBDAs black boxes? Should we avoid black boxes?

A: No. They are not black boxes and black boxes should not be avoided. SUM() is a black box as are all Excel functions because we cannot see their source. Even so, we use them because we know what they do and we know they work well. LAMBDAs, on the other hand, are not black boxes. We can see their formulas in Name Manager, or, more easily, in AFE where we can see imported LAMBDA documentation.

Q: Do LAMBDAs make models non-transparent?

A: If we mean not viewable from Excel's interface? No. LAMBDA formulas are viewable in Name Manager and AFE.

Q: Are LAMBDAs harder to understand?

A: That depends on our skill level and that is why LAMBDAs are so valuable. LAMBDAs can give those with less skill the power to do what those with great skill can do. If our skill level is high enough, we can read LAMBDAs just like normal Excel formulas because they are made from normal Excel formulas. And unlike normal Excel formulas, we can name them to convey intent. We can also add documentation IN the LAMBDA. If we use LET, we can break large formulas into smaller sections and name them. Named formulas, smaller named segments, and documentation greatly increase understandability.

Q: If we use a LAMBDA but do not understand how it works, how do we know it is okay to use?

A. We should test it.